\RequirePackage{fix-cm}
\documentclass[smallextended]{svjour3}       
\smartqed  
\usepackage{graphicx}
\usepackage{natbib}
\usepackage{amsmath,amsfonts,amssymb}
\usepackage{hyperref}
\usepackage{gensymb}

\newcommand{\TT}{\mathbf{T}}
\newcommand{\lo}{\mathrm{low}}
\newcommand{\up}{\mathrm{upp}}

\usepackage[usenames]{xcolor}  
\begin{document}

\title{Graphical tests of independence for general distributions\thanks{We acknowledge the financial support of Grant Agency of Czech Republic (Project No.\ 19-04412S)}}



\author{Ji\v{r}\'{\i} Dvo\v{r}\'{a}k         \and
        Tom\'{a}\v{s} Mrkvi\v{c}ka 
}


\institute{J. Dvo\v{r}\'{a}k \at
              Dept. of Probability and Mathematical Statistics, Faculty of Mathematics and Physics, Charles University, Sokolovsk\' a 83, 186 75 Prague, Czech Republic \\
              \email{dvorak@karlin.mff.cuni.cz}           
           \and
           T. Mrkvi\v{c}ka \at
               Dept. of Applied Mathematics and Informatics,  Faculty of Economics,  University of South Bohemia, Studentsk{\'a} 13,   37005 \v{C}esk\'e Bud\v{e}jovice, Czech Republic\\ 
         \email{mrkvicka.toma@gmail.com}  
}

\date{Received: date / Accepted: date}

\maketitle

\begin{abstract}
{We propose two model-free, permutation-based tests of independence between a pair of random variables. The tests can be applied to samples from any bivariate distribution: continuous, discrete or mixture of those, with light tails or heavy tails, \ldots The tests take advantage of the recent development of the global envelope tests in the context of spatial statistics. Apart from the broad applicability of the tests, their main benefit lies in the graphical interpretation of the test outcome: in case of rejection of the null hypothesis of independence, the combinations of quantiles in the two marginals are indicated for which the deviation from independence is significant. This information can be used to gain more insight into the properties of the observed data and as a guidance for proposing more complicated models and hypotheses. We assess the performance of the proposed tests in a simulation study and compare them to several well-established tests of independence. Furthermore, we illustrate the use of the tests and the interpretation of the test outcome in two real datasets consisting of meteorological reports (daily mean temperature and total daily precipitation, having an atomic component at 0 millimeters) and road accidents reports (type of road and the weather conditions, both variables having categorical distribution).}

\keywords{Independence \and Permutation test \and Visualization \and Empirical distribution function \and Intensity function}
\subclass{62G10 \and 62H15}
\end{abstract}

\section{Introduction}
\label{intro}
The problem of testing the null hypothesis of independence of two random variables is very important in statistics. There exist many methods to discover the possible dependence but all of them concentrate on summarizing the information over the whole distribution and as such cannot detect the combinations of quantiles where deviation from independence occurs. Our aim in this paper is to introduce tests of independence which are general, i.e. they can be used for arbitrary bivariate distributions \emph{without any assumptions}, and are graphical in nature, providing a graphical two-dimensional output where the combinations of quantiles causing the possible rejection are indicated.

Having two continuous distributions, an independence test usually computes a measure of association belonging to interval $[-1,1]$. Examples include the Pearson's correlation test which assumes normality and the Spearman's or Kendall's tests which are nonparametric and appropriate also for ordinal variables, see e.g. \citet{HollanderWolfe1973} for details. In our study we employ these tests using the implementation in the \texttt{R} package \texttt{stat} \citep{R2018}.

The Hoeffding's independence test measures the distance of the joint cumulative distribution function (CDF) to product of the marginal CDFs \citep{WildingMudholkar2008} by the Cram\' er-von Mises metric. The \texttt{Hmisc} \citep{Hmisc} package can be used to perform the test. The Genest's test \citep{GenestRemillard2004} uses the M\" obius decomposition of an empirical copula process and the distance is also measured by the Cram\' er-von Mises metric. The \texttt{copula} package \citep{copula} implements the test.

The permutation versions of all the tests mentioned above can be obtained by randomly permuting the order of one of the samples \citep{HajekEtal1999} to obtain Monte Carlo replicates of the data under the null hypothesis of independence and performing the test in a Monte Carlo fashion. The permutation version of the Hoeffding's test can be performed with the Cram\' er-von Mises metric but also with the supremal metric in the Kolmogorov-Smirnov sense. In our simulation study we employ the permutation test with Pearson's correlation coefficient as the test statistics and Hoeffding's test with both metrics. The supremal metric is principally different from other independence tests discussed above since it can also detect the deviation from independence for a single combination of quantiles.

Having two categorical distributions, the classical $\chi^2$ test in contingency tables is usually applied, possibly in the permutation version with the same test statistic. The $\chi^2$ test can be also applied to continuous distributions after discretization, but the choice of discretization is arbitrary, making the test highly dependent on the choices made by the user.

Having a general distribution, it is possible to apply the $Z$-transformation of the Pearson's correlation coefficient for which \citet{Hawkins1989} proved the asymptotic normality under the assumption of finite fourth moments. The asymptotic variance depends on the moments of the underlying distribution which makes the use of the test problematic in practice.

The description of more recent and more involved methods can be found in \citet{testforDEP}. These methods use different test statistics which again summarize the information along the whole distribution. As generally accepted, there does not exist an universally most powerful test, see e.g. \citet{testforDEP}. Therefore we compare the performance of our proposed methods only with the well-established methods, showing the proposed methods are competitive with them and delineating the settings where the proposed methods can be expected to perform well or perform poorly, respectively. Having performance comparable to other methods, the graphical interpretation is what makes our proposed methods an interesting choice for testing independence.

{Concerning graphical methods,} \citet{FisherSwitzer1985} describe the so-called $\chi$-plot and \citet{GenestBoies2012} describe the so-called Kendall's plot for detecting dependence in a bivariate sample. These plots, even though they are able to detect a mixture of positive and negative correlation, have to be taken as an exploratory tool only. On the other hand, our proposed methods provide both a visual output and a formal outcome in the form of $p$-value. 

Recently, the global envelope test \citep{MyllymakiEtal2017, MrkvickaEtal2017} was introduced for solving the multiple comparison problem in a graphical way. Specifically, if the multivariate test statistic $\mathbf T_0$ is observed together with its Monte Carlo replications $\mathbf T_1, \ldots , \mathbf T_s$ (e.g. simulations from the null model), then the global envelope test produces the envelope $(\mathbf T^{\rm low}, \mathbf T^{\rm up})$ such that $\mathbf T_0$ is outside of the envelope in at least one element of $\mathbf T_0$ with prescribed probability $\alpha$ under the null hypothesis. {The decision rule is then the following: \emph{reject} the null hypothesis if $\mathbf T_0$ leaves the envelope at any coordinate, otherwise \emph{not reject}.} The test is performed with family-wise error rate control.

Since the global envelope test can be applied to any multivariate test statistic, it can be applied to the sample CDF $\hat F(x,y)$ computed in a grid of $(x,y)$ values, {with the values formally arranged to a long vector}. This allows performing a permutation-based independence test which is able to detect {which arguments} of $\hat F(x,y)$ are responsible for the possible rejection. The test is applicable to any kind of distribution without any assumptions. In case of categorical distribution the test reveals which cells are responsible for rejection. It would be possible to sort the {permuted samples} by various other measures and obtain a $p$-value by a Monte Carlo test but we are interested only in measures which provide localized information about the causes of possible rejection. 

In this paper, we propose two test statistics for the permutation-based independence test. The first one is the above-mentioned sample CDF $\hat F(x,y)$ computed on a fine grid of quantiles. The second test statistic is constructed as follows. Assume we have a sample $\mathbb X= \{(x_1, y_1), \ldots , (x_n, y_n)\}$ from a two-dimensional distribution. The two-dimensional quantile-quantile plot (2D qq-plot) of $\mathbb X$ can be considered a regular homogeneous point process in the unit square $[0,1]^2$ under independence of the random variables {\citep{IllianEtal2008,MollerWaagepetersen2003,spatstat2015}}. The kernel estimator of the intensity function of the 2D qq-plot $\hat g(x,y)$ computed on a fine grid of quantiles is the second proposed test statistic. These tests can be used to reveal the information where (i.e. for which quantiles) the possible dependence appears for continuous distributions. For categorical distributions, {an adapted version of these tests} reveals which combinations of categories cause the possible dependence.

The supremal metric used in Hoeffding's test can also identify local deviations from independence, similarly to our proposed methods. {In a certain sense, the supremal metric forms an envelope with constant width}. The global envelope chosen here is instead based on the {pointwise} ranks and thus reflects possibly varying variability in different locations. Further, due to its nonparametric nature, it gives the same weight to every coordinate of the chosen test statistic. Another difference can be seen in the test statistic: whereas the Hoeffding's test statistic, just as the CDF, expresses the local deviations from independence in a cumulative way, the 2D qq-plot expresses it in a non-cumulative way. As a consequence, it is much easier to interpret the results. All the other tests of independence considered here are based on characteristics summarizing the whole distribution to a scalar test statistic.

The paper is organized as follows. First, we briefly describe the {principle of Monte Carlo testing and the} global envelope test and its features in Section~\ref{se:MC}. Then we describe in detail how these are applied to the independence testing in Section~\ref{se:TS}. In Section~\ref{se:SS} we investigate the performance of the proposed tests for several continuous distributions and compare them to the widely used tests of independence. We show that our tests match the nominal significance level and that in simple situations they can compete with other tests in terms of power. We also show situations where the performance of {one of} our tests is much better due to its local, non-cumulative nature. Further, we illustrate the use of our tests and their graphical interpretation in an example with continuous distribution with an atom in Section \ref{se:EA} and in an example with categorical distributions in Section \ref{se:EC}. The \texttt{R} code providing the implementation of the proposed tests is available for download at \url{http://msekce.karlin.mff.cuni.cz/~dvorak/software.html}.

\section{Monte Carlo testing}\label{se:MC}

\subsection{{General case}}
Let $\mathbf{T_0}$ be an observed test statistic. The principle of Monte Carlo testing lies in simulating the test statistic values under the assumption of validity of the tested null model \citep{DavisonHinkley1997}. Thus let 
$\mathbf{T}_1, \ldots , \mathbf{T}_s$ be such simulated test statistic values. The Monte Carlo $p$-value is 
\begin{align*}
    p=\frac{1}{s+1}\sum_{i=0}^s \mathbf 1(\mathbf{T}_i \prec \mathbf{T}_0),
\end{align*}
where $\mathbf 1$ is the indicator function and where $\mathbf{T}_i \prec \mathbf{T}_0$ indicates that $\mathbf{T}_i$ is {more extreme than (or same as)} $\mathbf{T}_0$ in a certain ordering. 

For univariate test statistic a natural ordering exists but for multivariate test statistic there exist many ways to define an ordering. In the following subsection we will shortly describe the recently defined multivariate ordering \citet{MyllymakiEtal2017,MrkvickaEtal2017,MyllymakiMrkvicka2020} which meets the desired intrinsic graphical interpretation (IGI) {defined below}.

\subsection{Global envelope test} \label{se:GET}
Consider now a $d$-dimensional test statistics $\TT_i=(T_{i1},\dots,T_{id})$, $i=0,\dots,s$. 
The proposed ordering is based on the pointwise ranks $R_{ik}${, constructed below,}
of the element $T_{ik}$ among the corresponding elements $T_{0k}, T_{1k}, \dots, T_{s k}$ of the $s+1$ vectors such that the lowest ranks correspond to the most extreme values of the statistics.
Let $r_{0k}, r_{1k}, \dots, r_{s k}$ be the raw ranks of $T_{0k}, T_{1k}, \dots, T_{s k}$, such that the smallest value has rank 1 {and the largest value is assigned rank $s+1$}. In the case of ties, the raw ranks are averaged. The two-sided pointwise ranks are then calculated as
\begin{align*}
  R_{ik}=
   \min(r_{ik}, s+1-r_{ik}).
\end{align*}

The extreme rank length (ERL) measure defined in \citet{MyllymakiEtal2017} induces the desired ordering. Roughly speaking, it ranks the vectors according to the number of the most extremes elements. Formally, the ERL measure of $\TT_i$ is defined based on the vector of the pointwise ordered ranks
$\mathbf{R}_i=(R_{i[1]}, R_{i[2]}, \dots , R_{i[d]})$, where the ranks are arranged from smallest to largest, i.e., $R_{i[k]} \leq R_{i[k^\prime]}$ whenever $k \leq k^\prime$. 
The ERL measure of $\TT_i$ is defined as
\begin{equation}
\label{eq:lexicalrank}
   E_i = \frac{1}{s+1}\sum_{i^\prime=0}^{s} \mathbf{1} (\mathbf{R}_{i^\prime} \prec \mathbf{R}_i)
\end{equation}
where
\[
\mathbf{R}_{i^\prime} \prec \mathbf{R}_{i} \quad \Longleftrightarrow\quad
  \exists\, n\leq d: R_{i^\prime[k]} = R_{i[k]} \forall\, k < n, \; {\mathrm{ and }} \; R_{i^\prime[n]} < R_{i[n]}.
\]
The division by $s+1$ leads to normalized ranks that attain values between 0 and 1. Consequently, the ERL measure corresponds to the extremal depth of \citet{NarisettyNair2016}.

The probability of having a tie in the ERL measure is rather small {in practical scenarios}, thus the ERL effectively solves the ties problem which often appears in  ordering of multivariate vectors using ranks. The final $p$~value of the Monte Carlo test is $$p_\text{erl}=  \sum_{i=0}^{s} \frac{1}{s+1} \mathbf 1 (E_i \leq E_0).$$

Let $e_\alpha\in \mathbb{R}$ be the largest of the $E_i$ such that the number of those $i$ for which $E_i< e_{(\alpha)}$
is less or equal to $\alpha s$. Further let $I_\alpha = \{i\in 0,\dots, s: E_i \geq e_{(\alpha)} \}$ be the index set of vectors less or as extreme as $e_\alpha$. 
Then the $100(1-\alpha)$\% global ERL envelope induced by $E_i$ is
\begin{align}\label{erl_envelope}
\TT^{(\alpha)}_{\lo \, k}= \underset{i \in I_\alpha}{{\min}}\ T_{i k}
\quad\text{and}\quad
\TT^{(\alpha)}_{\up \, k}= \underset{i \in I_\alpha}{{\max}}\ T_{i k} \quad \text{for } k = 1, \ldots , d,
\end{align}
see \citet{NarisettyNair2016} and \citet{MrkvickaEtal2020}.

The $100(1-\alpha)$\% global ERL envelope $[\TT_{\lo\,k}^{(\alpha)}, \TT_{\up\,k}^{(\alpha)}], {k = 1, \ldots, d,}$ has an important {\it intrinsic graphical interpretation} (IGI) property, i.e.  
\begin{enumerate}
 \item $T_{ik} < \TT_{\lo\,k}^{(\alpha)}$ or $T_{ik} > \TT_{\up\,k}^{(\alpha)}$ for some $k = 1, \ldots , d$
{$\iff$} $E_i<e_{(\alpha)}$ for every $i=0, \ldots , s$;
 \item $\TT_{\lo\,k}^{(\alpha)} \leq T_{ik} \leq \TT_{\up\,k}^{(\alpha)}$ for all $k = 1, \ldots , d$ {$\iff$} $E_i\geq e_{(\alpha)}$ for every $i=0, \ldots , s$.
\end{enumerate}

\section{{Proposed independence tests}}
\label{se:TS}
Consider now a bivariate random vector $\mathbf X = (X,Y)$ and the null hypothesis of independence of $X$ and $Y$ which is about to be tested. Assume that a sample  $\mathbb X= \{(x_1, y_1), \ldots , (x_n, y_n)\}$ is observed.

The first test statistic being considered in connection with the global envelope test described above is the sample cumulative distribution function $\hat F(x,y)$ computed on a fine of grid points $(u,v)_1, \ldots , (u,v)_d$. The grid points here are two-dimensional but it does not play any role in global envelope testing, since due to the ranking nature, every point is taken with the same weight and any correlation between the elements of the test statistic is treated by resampling of the whole vector. This test statistic is of cumulative nature 
and as such it is sensitive to small departures from the null model which accumulate into a possibly significant departure.

In order to define a local test {with easier interpretation} we define the following test statistic. It is the kernel estimate of the intensity {function} $\hat g(x,y)$ of the 2D qq-plot, constructed as follows. Let $\mathbb Q$ denote the 2D point pattern $\mathbb Q = \{(q_1^1,q_1^2), \ldots , (q_n^1,q_n^2)\}$, where $q_i^1$ {is the sample quantile corresponding to the value $x_i$ in the sample} $x_1, \ldots , x_n$, and similarly $q_i^2$ {is the sample quantile corresponding to the value $y_i$ in the sample} $y_1, \ldots , y_n$. $\mathbb Q$ is a point process in the observation window $[0,1]^2$. The intensity $g(x,y)$ of the point pattern $\mathbb Q$ is defined by 
\begin{align*}
    \mathbb E N(A) = \int_A g(x,y) {\lambda_2(\mathrm{d}x,\mathrm{d}y)},
\end{align*}
where $\lambda_2$ denotes the two-dimensional Lebesgue measure and $N(A)$ denotes the number of points {of $\mathbb{Q}$} in a Borel set $A\subset [0,1]^2$.
Constant intensity of $\mathbb Q$ indicates independence whereas accumulation of points in a certain area indicates steeper increase of the CDF values at the given combination of quantiles, meaning observations with such combination of values are more likely to appear than what is expected under independence. On the other hand, absence of points in a certain area indicates less steep increase of the CDF values at the given combination of quantiles, meaning observations with such combination of values are less likely to appear than what is expected under independence.

The common way to estimate the intensity of a point process \citep{IllianEtal2008} is to use a nonparametric kernel estimator 
\begin{align*}
    \hat g(x,y) = \frac{1}{e(x,y)} \sum_{i=1}^n \mathbf k_\sigma((x,y)-(x_i,y_i)),
\end{align*}
where $\mathbf k_\sigma$ is a probability density function usually called kernel function with bandwidth $\sigma$, $e(x,y)=\int_{[0,1]^2} \mathbf k_\sigma((x,y)-(x',y')){\lambda_2(\mathrm{d}x,\mathrm{d}y)}$ is the correction for bias due to edge effects {The edge effects, often appearing in statistics for point processes, are in this context closely related to the problem of bounded support in kernel estimation of a probability density function in classical statistics}.

The common choice of kernel function is the isotropic Gaussian probability density function with standard deviation $\sigma$. The choice of the bandwidth parameter $\sigma$ involves a trade-off between bias and variance: as bandwidth increases, typically the bias increases and variance decreases. To avoid the {need of manual} bandwidth selection, we propose here to use the Scott's rule of thumb, which is in two dimensions equivalent to Silverman's rule of thumb \citep{Silverman1986}. Hence we suggest to use $\sigma = n^{-1/6} \cdot \sqrt{1/12}$, where $1/12$ is the limit variance of $q_1^1, \ldots , q_n^1$ for $n\longrightarrow \infty$, i.e. the limit variance of $n$ regularly spaced points in the interval [0,1].

{Note that the proposed test statistic is defined as a two-dimensional kernel estimate and hence it is appropriate for continuous distributions for which the probability of ties appearing in any coordinate of the observed data points is 0. If an atom appears in the distribution of one of the marginals, a one-dimensional analogue of the point pattern intensity estimate can be used, replacing the appropriate part of the $\hat g(x,y)$ surface. Recall here that the global envelope test assigns the same weight to every grid point, therefore we use the one-dimensional estimate for all grid rows (or columns) which correspond to the quantiles covered by the atom. The data example in Section~\ref{se:EA} provides an illustration, see in particular Figure~\ref{fig:weather_qq}.}

{If atoms appear in the distribution of both marginals, the appropriate part of the $\hat g(x,y)$ surface can be replaced by the constant value corresponding to the number of observations attaining the given combination of values. Again, to achieve proper weighting in the global envelope test, we use these zero-dimensional estimates for all grid points which correspond to the quantiles covered by the combination of atoms in the two marginals.}

{We propose the one-dimensional and zero-dimensional estimates be scaled so that the desired property $\int_0^1 \int_0^1 \hat g(x,y) \, \mathrm{d}x \, \mathrm{d} y = 1$, which holds for the two-dimensional kernel estimator of the intensity function, holds also in this setting. However, this is not necessary for the testing since the global envelope test is based on ranks only and such scaling does not affect the outcome of the test.}

If the distributions are purely categorical the situation simplifies and it is possible to define the test statistic $\hat c = (\hat c_{11}, \ldots, \hat c_{k_1k_2}) $ as the vector of all $k_1k_2$ elements of the appropriate contingency table and apply the global envelope test on that statistic. This approach is applicable also for 3 or more dimensional contingency tables, however, in that case the permutation scheme needs to be more delicate and related to the precise formulation of the null hypothesis of independence.

When the test statistic is defined and computed for the data, the Monte Carlo test requires to compute the same test statistic for data replicated under the null hypothesis. This is achieved in the same way as in the classical permutation independence tests \citep{HajekEtal1999} by permuting the order of $\{ y_1, \ldots y_n \}$. This permutation scheme secures exchangeability of all tests statistics \citep{Kallenberg2006} computed from sample $\{(x_1, \pi (y_1)), \ldots , (x_n, \pi (y_n))\}$, where $\pi$ is the permutation, therefore any such permutation test will achieve the prescribed significance level $\alpha$.

\section{Simulation study}
\label{se:SS}

In all the experiments described below we perform the tests on significance level 0.01, with different sample sizes $n$. In each experiment we generate $5\,000$ independent samples and report the rejection rate for the individual tests. For permutation-based tests 999 random permutations are used. {Where needed, the empirical distribution function is computed on the $20 \times 20$ grid of points, the coordinates of which are determined as 5\%, 10\%, \ldots  empirical quantiles in each coordinate. The test statistic $\hat g(x,y)$ is computed on a regular grid of $32 \times 32$ points. These choices were made based on the results of pilot studies, not reported here. However, the performance of the proposed tests is rather robust with respect to these choices.}

{Apart from the proposed tests, the tests considered in this study are well-known and commonly used, i.e. the tests based on the Pearson's (denoted Pea in the following), Spearmans's (Spe), Kendall's (Ken) correlation coefficient and the Hoeffding's (Hoe) and the Genest's (Gen) independence tests. We further consider the permutation tests with Hoeffding's test statistic and supremum metric (DevS) and with Cram\' er von Mises metric (DevI) and finally the permutation test with Pearson's correlation coefficient (PeaP). These are accompanied by the permutation tests with the proposed tests statistic $\hat F(x,y)$ (CDF) and $\hat g(x,y)$ (QQ).}

\subsection{Size of the tests}
We first investigate whether the tests considered in this paper match the nominal significance level. We employ two different scenarios where the components are independent and have 1) standard normal distribution and 2) Pareto(4) distribution. The normal distribution is chosen as a benchmark distribution which is very well understood and for which all the tests are expected to perform well. The Pareto distribution is chosen in order to cover the case with heavier tails.

The top part of Table~\ref{tab:H0} gives the rejection rates for the case of the standard normal distribution. As expected, all the tests match the nominal significance level 0.01 very closely. The bottom part of Table~\ref{tab:H0} shows the results for the Pareto distribution. Again, all the tests (except the test based on the Pearson's correlation coefficient) match the nominal significance level very closely.

\begin{table}[t]
\centering
\begin{tabular}{c|cccccccccc}
  \hline
$n$ & Pea   & Spe   & Ken   & Hoe   & Gen   & CDF   & QQ    & DevS  & DevI  & PeaP \\ \hline
50  & 0.009 & 0.010 & 0.008 & 0.013 & 0.011 & 0.008 & 0.011 & 0.012 & 0.012 & 0.011 \\ 
100 & 0.007 & 0.009 & 0.008 & 0.012 & 0.010 & 0.009 & 0.008 & 0.010 & 0.010 & 0.010 \\ 
150 & 0.009 & 0.010 & 0.009 & 0.011 & 0.009 & 0.009 & 0.009 & 0.009 & 0.010 & 0.010 \\
200 & 0.009 & 0.011 & 0.011 & 0.011 & 0.010 & 0.009 & 0.010 & 0.009 & 0.011 & 0.012 \\ 
250 & 0.010 & 0.009 & 0.008 & 0.009 & 0.009 & 0.008 & 0.009 & 0.009 & 0.009 & 0.012 \\ 
300 & 0.010 & 0.009 & 0.009 & 0.010 & 0.010 & 0.007 & 0.007 & 0.012 & 0.010 & 0.009 \\ \hline \hline
$n$ & Pea   & Spe   & Ken   & Hoe   & Gen   & CDF   & QQ    & DevS  & DevI  & PeaP \\ \hline
50  & 0.019 & 0.012 & 0.012 & 0.016 & 0.012 & 0.009 & 0.009 & 0.009 & 0.012 & 0.008 \\ 
100 & 0.020 & 0.009 & 0.009 & 0.012 & 0.010 & 0.008 & 0.010 & 0.008 & 0.010 & 0.011 \\ 
150 & 0.021 & 0.009 & 0.009 & 0.011 & 0.009 & 0.008 & 0.009 & 0.010 & 0.010 & 0.010 \\ 
200 & 0.017 & 0.010 & 0.010 & 0.012 & 0.012 & 0.007 & 0.009 & 0.012 & 0.012 & 0.011 \\  
250 & 0.019 & 0.010 & 0.010 & 0.011 & 0.010 & 0.009 & 0.008 & 0.009 & 0.009 & 0.010 \\ 
300 & 0.017 & 0.009 & 0.009 & 0.008 & 0.009 & 0.008 & 0.008 & 0.010 & 0.011 & 0.009 \\ \hline
\end{tabular}
\caption{Simulations under the null hypothesis of independence, marginal distributions are standard normal (top part) or Pareto (bottom part). The sample size is $n$. The fraction of rejections from $5\,000$ replications is reported for the individual tests.}\label{tab:H0}
\end{table}

\subsection{Power of the tests}
\begin{figure}
    \centering
    \includegraphics[width=\textwidth]{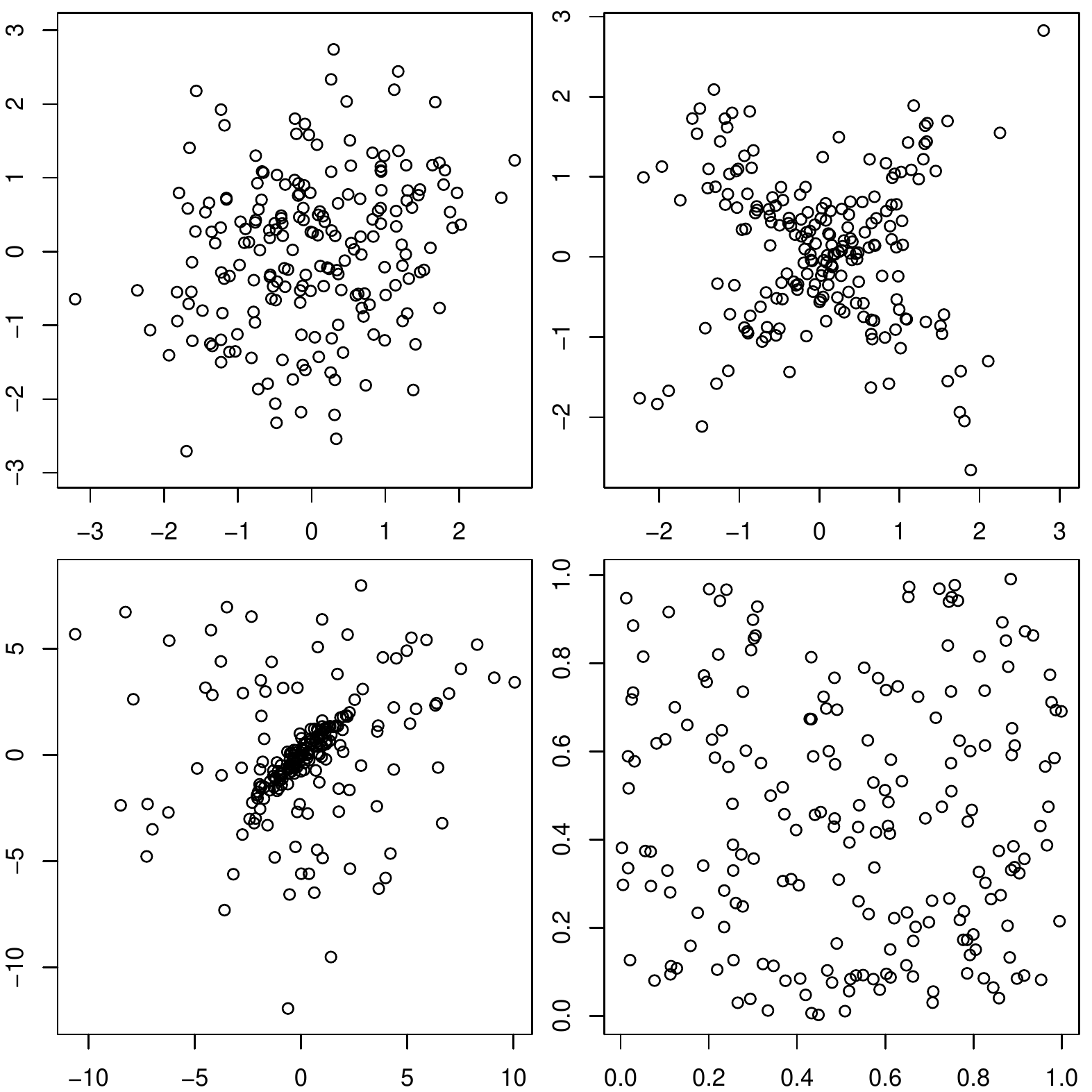}
    \caption{Examples of realizations for different scenarios considered in the simulation study. Top left: experiment 1 with $\rho=0.3$, top right: experiment 2 with $\rho=0.9$, bottom left: experiment 3 with $\rho=0.9$, bottom right: experiment 4.}
    \label{fig:realizations_H1}
\end{figure}

To investigate the power of the various tests discussed in this paper we consider four different scenarios given below. They cover the cases where the theoretical (Pearson) correlation between the marginals is either zero or non-zero and the cases where the deviations from independence have either a localized nature or appear over the whole range of data values. Figure~\ref{fig:realizations_H1} shows examples of realizations corresponding to the Experiments 1 to 4 below.

\subsubsection{Experiment 1}
First we consider the well-understood case of bivariate normal distribution with correlation $\rho$ (and zero mean and unit variance) for the choices $\rho=0.1, 0.2, 0.3$ and different sample sizes.

The fractions of rejections of the different tests are shown in Table~\ref{tab:H1_Exp1}. The tests based on correlation coefficients (Pea, Spe, Ken) perform the best, also in the permutation version (PeaP). The Hoeffding and Genest tests have also very high power. The CDF-based tests (CDF, DevS, DevI) have somewhat smaller power but the DevI test is almost competitive with the tests based on correlation coefficients. This can be explained by noting that in this model the deviation from independence is present over the whole range of data values which gradually accumulates until it is recognized as statistically significant. On the other hand, the intensity based test (QQ) performs slightly worse here because no part of the distribution domain exhibits strong departure from independence.

\begin{table}[t]
\centering
\begin{tabular}{c|cccccccccc}
  \hline
$n$ & Pea   & Spe   & Ken   & Hoe   & Gen   & CDF   & QQ    & DevS  & DevI  & PeaP \\ \hline
50  & 0.029 & 0.025 & 0.023 & 0.029 & 0.022 & 0.020 & 0.015 & 0.035 & 0.030 & 0.034 \\
100 & 0.057 & 0.056 & 0.055 & 0.055 & 0.050 & 0.038 & 0.021 & 0.051 & 0.047 & 0.058 \\ 
150 & 0.082 & 0.074 & 0.073 & 0.070 & 0.062 & 0.050 & 0.025 & 0.072 & 0.073 & 0.095 \\ 
200 & 0.127 & 0.114 & 0.110 & 0.101 & 0.093 & 0.066 & 0.033 & 0.089 & 0.096 & 0.133 \\ 
250 & 0.158 & 0.138 & 0.137 & 0.124 & 0.116 & 0.081 & 0.046 & 0.111 & 0.131 & 0.172 \\ 
300 & 0.198 & 0.174 & 0.173 & 0.157 & 0.147 & 0.105 & 0.048 & 0.133 & 0.161 & 0.203 \\ \hline \hline
$n$ & Pea   & Spe   & Ken   & Hoe   & Gen   & CDF   & QQ    & DevS  & DevI  & PeaP \\ \hline
50  & 0.116 & 0.101 & 0.101 & 0.106 & 0.088 & 0.064 & 0.033 & 0.093 & 0.098 & 0.129 \\ 
100 & 0.285 & 0.245 & 0.245 & 0.225 & 0.213 & 0.142 & 0.075 & 0.178 & 0.228 & 0.291 \\ 
150 & 0.447 & 0.398 & 0.395 & 0.357 & 0.339 & 0.246 & 0.127 & 0.290 & 0.392 & 0.473 \\ 
200 & 0.607 & 0.556 & 0.554 & 0.499 & 0.492 & 0.364 & 0.204 & 0.383 & 0.529 & 0.617 \\ 
250 & 0.722 & 0.672 & 0.670 & 0.613 & 0.599 & 0.472 & 0.273 & 0.488 & 0.644 & 0.739 \\ 
300 & 0.822 & 0.771 & 0.772 & 0.715 & 0.703 & 0.576 & 0.353 & 0.582 & 0.768 & 0.833 \\ \hline \hline
$n$ & Pea   & Spe   & Ken   & Hoe   & Gen   & CDF   & QQ    & DevS  & DevI  & PeaP \\ \hline
50  & 0.328 & 0.281 & 0.273 & 0.270 & 0.239 & 0.163 & 0.093 & 0.214 & 0.266 & 0.344 \\ 
100 & 0.689 & 0.618 & 0.619 & 0.573 & 0.546 & 0.407 & 0.232 & 0.438 & 0.592 & 0.695 \\ 
150 & 0.885 & 0.836 & 0.837 & 0.789 & 0.779 & 0.659 & 0.429 & 0.651 & 0.822 & 0.883 \\ 
200 & 0.965 & 0.941 & 0.941 & 0.913 & 0.906 & 0.817 & 0.615 & 0.804 & 0.929 & 0.965 \\  
250 & 0.990 & 0.979 & 0.978 & 0.965 & 0.962 & 0.914 & 0.761 & 0.885 & 0.968 & 0.989 \\ 
300 & 0.998 & 0.994 & 0.994 & 0.989 & 0.987 & 0.964 & 0.856 & 0.947 & 0.995 & 0.998 \\ \hline
\end{tabular}
\caption{Experiment 1, bivariate normal distribution with correlation $\rho$ for choices $\rho=0.1$ (top part), $\rho=0.2$ (middle part) and $\rho=0.3$ (bottom part). The sample size is $n$. The fraction of rejections from $5\,000$ replications is reported for the individual tests.}\label{tab:H1_Exp1}
\end{table}

\subsubsection{Experiment 2}
In this experiment we consider the mixture (with equal weights) of two centered, unit variance bivariate normal distributions, one of them having correlation $\rho$ and the other having correlation $-\rho$. We investigate two different values, $\rho = 0.7$ and $\rho = 0.9$. In this setting the theoretical (Pearson) correlation between the marginals is 0 and the data points form a kind of cross rather than a cloud (see the top right part of Figure~\ref{fig:realizations_H1}). Furthermore, the deviation from independence appears over the whole range of data values.

Table~\ref{tab:H1_Exp2} shows the performance of the different tests. The tests based on correlation coefficients (Pea, Spe, Ken, PeaP) have low power which does not grow with the increasing sample size but grows with increasing value of $\rho$. Most of the permutation-based test (CDF, QQ, DevI) have very good power even for the smaller value of $\rho$. The exception is the DevS test which has only very low power in this case. {This can be attributed to the deviations from independence occurring in the periphery of the distribution where the joint CDF is necessarily close to the product of the marginal CDFs and the supremum of the difference is often realized in the center of the distribution, completely overlooking the peripheral parts.} The Hoeffding and Genest tests have rather small power for smaller value of $\rho$ and smaller sample sizes but it gets higher with increasing sample size and increasing $\rho$. The highest power is observed for the QQ test, followed by the CDF test and the DevI test.

\begin{table}[t]
\centering
\begin{tabular}{c|cccccccccc}
  \hline
$n$ & Pea   & Spe   & Ken   & Hoe   & Gen   & CDF   & QQ    & DevS  & DevI  & PeaP \\ \hline
50  & 0.066 & 0.027 & 0.037 & 0.041 & 0.022 & 0.035 & 0.057 & 0.015 & 0.055 & 0.070 \\ 
100 & 0.062 & 0.025 & 0.033 & 0.039 & 0.026 & 0.089 & 0.169 & 0.020 & 0.082 & 0.062 \\ 
150 & 0.068 & 0.024 & 0.035 & 0.050 & 0.034 & 0.146 & 0.361 & 0.024 & 0.146 & 0.066 \\ 
200 & 0.070 & 0.026 & 0.039 & 0.066 & 0.046 & 0.250 & 0.549 & 0.031 & 0.233 & 0.066 \\ 
250 & 0.067 & 0.027 & 0.039 & 0.073 & 0.055 & 0.359 & 0.722 & 0.034 & 0.369 & 0.065 \\ 
300 & 0.072 & 0.026 & 0.037 & 0.086 & 0.066 & 0.497 & 0.850 & 0.042 & 0.509 & 0.068 \\ \hline \hline
$n$ & Pea   & Spe   & Ken   & Hoe   & Gen   & CDF   & QQ    & DevS  & DevI  & PeaP \\ \hline
50  & 0.115 & 0.045 & 0.072 & 0.112 & 0.061 & 0.166 & 0.687 & 0.041 & 0.189 & 0.109 \\ 
100 & 0.105 & 0.039 & 0.065 & 0.186 & 0.113 & 0.527 & 0.993 & 0.086 & 0.555 & 0.114 \\ 
150 & 0.116 & 0.045 & 0.075 & 0.406 & 0.265 & 0.872 & 1.000 & 0.164 & 0.940 & 0.106 \\ 
200 & 0.115 & 0.042 & 0.073 & 0.750 & 0.565 & 0.993 & 1.000 & 0.268 & 0.998 & 0.107 \\ 
250 & 0.115 & 0.043 & 0.078 & 0.970 & 0.893 & 1.000 & 1.000 & 0.378 & 1.000 & 0.112 \\ 
300 & 0.116 & 0.042 & 0.071 & 0.998 & 0.993 & 1.000 & 1.000 & 0.486 & 1.000 & 0.110 \\ \hline
\end{tabular}
\caption{Experiment 2, mixture of two bivariate normal distributions with correlation $\rho$ and $-\rho$ for choices $\rho=0.7$ (top part) and $\rho=0.9$ (bottom part). The sample size is $n$. The fraction of rejections from $5\,000$ replications is reported for the individual tests.}\label{tab:H1_Exp2}
\end{table}

\subsubsection{Experiment 3}
Now we investigate the mixture (with equal weights) of two centered bivariate normal distributions. The first one has independent components with variance 16, the second one has unit variance and correlation $\rho$ for choices $\rho=0.8$ and $\rho=0.9$. See the bottom left part of Figure~\ref{fig:realizations_H1} for a simulated realization.

The deviation from independence in this model is localized in the center rather than spread across the whole range of data values. This causes the QQ and DevS tests to have very high power here, see Table~\ref{tab:H1_Exp3}. In fact, the QQ test is the most powerful from all the tests considered here. We further observe that the Hoeffding test has almost as high power, followed by the Genest test, DevI, Ken and Spe tests. Tests based on the Pearson's correlation coefficient perform poorly in this scenario.

\begin{table}[t]
\centering
\begin{tabular}{c|cccccccccc}
  \hline
$n$ & Pea   & Spe   & Ken   & Hoe   & Gen   & CDF   & QQ    & DevS  & DevI  & PeaP \\ \hline
50  & 0.068 & 0.137 & 0.231 & 0.459 & 0.237 & 0.236 & 0.693 & 0.369 & 0.211 & 0.064 \\ 
100 & 0.072 & 0.246 & 0.450 & 0.814 & 0.577 & 0.580 & 0.990 & 0.716 & 0.515 & 0.080 \\ 
150 & 0.079 & 0.347 & 0.614 & 0.957 & 0.841 & 0.842 & 1.000 & 0.913 & 0.804 & 0.088 \\  
200 & 0.083 & 0.464 & 0.749 & 0.991 & 0.961 & 0.957 & 1.000 & 0.976 & 0.953 & 0.094 \\ 
250 & 0.098 & 0.562 & 0.837 & 0.999 & 0.995 & 0.993 & 1.000 & 0.993 & 0.990 & 0.096 \\ 
300 & 0.098 & 0.655 & 0.897 & 1.000 & 0.999 & 0.999 & 1.000 & 0.998 & 0.998 & 0.101 \\ \hline \hline
$n$ & Pea   & Spe   & Ken   & Hoe   & Gen   & CDF   & QQ    & DevS  & DevI  & PeaP \\ \hline
50  & 0.071 & 0.170 & 0.318 & 0.651 & 0.325 & 0.342 & 0.885 & 0.515 & 0.255 & 0.065 \\ 
100 & 0.077 & 0.313 & 0.590 & 0.944 & 0.742 & 0.762 & 1.000 & 0.875 & 0.628 & 0.083 \\ 
150 & 0.085 & 0.446 & 0.764 & 0.995 & 0.951 & 0.958 & 1.000 & 0.981 & 0.892 & 0.087 \\ 
200 & 0.091 & 0.575 & 0.881 & 1.000 & 0.994 & 0.993 & 1.000 & 0.997 & 0.984 & 0.098 \\ 
250 & 0.108 & 0.682 & 0.940 & 1.000 & 1.000 & 1.000 & 1.000 & 1.000 & 0.997 & 0.113 \\ 
300 & 0.111 & 0.763 & 0.970 & 1.000 & 1.000 & 1.000 & 1.000 & 1.000 & 1.000 & 0.119 \\ \hline
\end{tabular}
\caption{Experiment 3, mixture of two bivariate normal distributions, one with independent components and large variance, the other one with small variance and correlation $\rho$ for choices $\rho=0.8$ (top part) and $\rho=0.9$ (bottom part). The sample size is $n$. The fraction of rejections from $5\,000$ replications is reported for the individual tests.}\label{tab:H1_Exp3}
\end{table}

\subsubsection{Experiment 4}
In the last experiment we consider a distorted uniform distribution on the square $[0,1]^2$ with the probability density function $f(x,y)=1$ except that $f(x,y)=0$ for $x \in (0.35,0.65), y \in (0.85,1)$ and $f(x,y)=2$ for $x \in (0.35,0.65)$, $y \in (0,0.15)$. It can be shown by direct calculation that for this distribution the (theoretical) Pearson's correlation coefficient is 0 and hence the tests based on correlation coefficients are expected to perform poorly in this setting. On the other hand, the deviation from independence (uniform density in this case) is localized and the QQ test are expected to have high power.

Table~\ref{tab:H1_Exp4} shows the fractions of rejection for various tests and different sample sizes. Note that in this experiments the considered sample sizes are higher than in the previous experiment in order to see the increasing power of (some of) the tests. Indeed, the tests based on the correlation coefficients (Pea, Spe, Ken, PeaP) have no power at all. The Hoeffding, Genest and DevS, DevI tests exhibit only small power which grows slowly with increasing sample size. The power of the CDF test is somewhat higher and, as expected, the intensity based test (QQ) performs the best in this setting. This can be attributed to the localized nature of the departure from independence in the model.

\begin{table}[t]
\centering
\begin{tabular}{c|cccccccccc}
  \hline
$n$ & Pea   & Spe   & Ken   & Hoe   & Gen   & CDF   & QQ    & DevS  & DevI  & PeaP \\ \hline
100 & 0.010 & 0.011 & 0.011 & 0.020 & 0.018 & 0.025 & 0.037 & 0.025 & 0.021 & 0.013 \\ 
200 & 0.009 & 0.010 & 0.009 & 0.034 & 0.028 & 0.058 & 0.118 & 0.041 & 0.024 & 0.012 \\ 
300 & 0.011 & 0.010 & 0.010 & 0.052 & 0.050 & 0.098 & 0.236 & 0.060 & 0.046 & 0.014 \\  
400 & 0.012 & 0.012 & 0.012 & 0.076 & 0.073 & 0.155 & 0.401 & 0.085 & 0.075 & 0.013 \\ 
500 & 0.011 & 0.011 & 0.011 & 0.120 & 0.117 & 0.230 & 0.545 & 0.115 & 0.121 & 0.011 \\ \hline
\end{tabular}
\caption{Experiment 4, distorted uniform distribution on the square $[0,1]^2$. The sample size is $n$. The fraction of rejections from $5\,000$ replications is reported for the individual tests.}\label{tab:H1_Exp4}
\end{table}

\section{Example for continuous distributions with an atom}
\label{se:EA}

\begin{figure}[t]
    \centering
    \includegraphics[width=\textwidth]{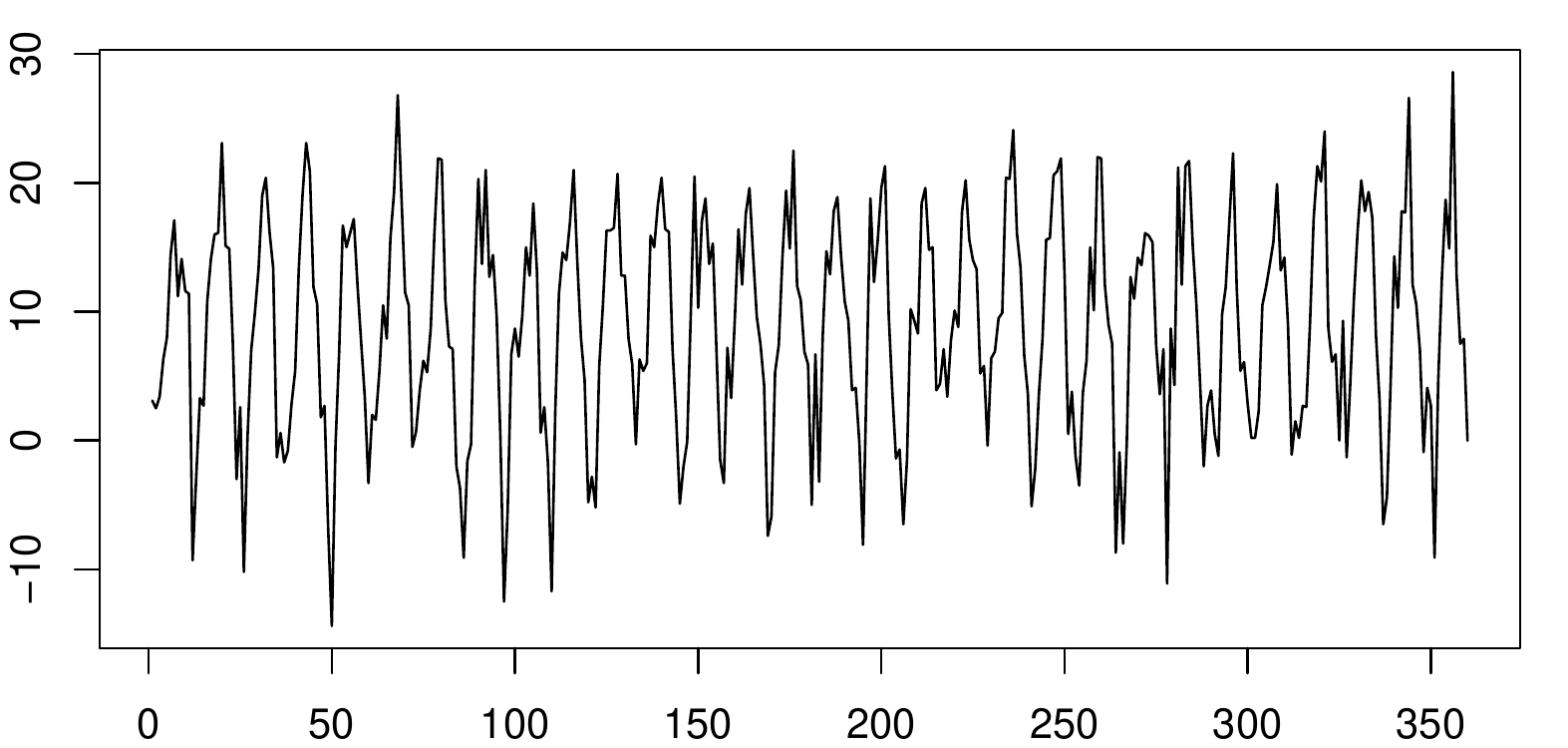}
    \includegraphics[width=\textwidth]{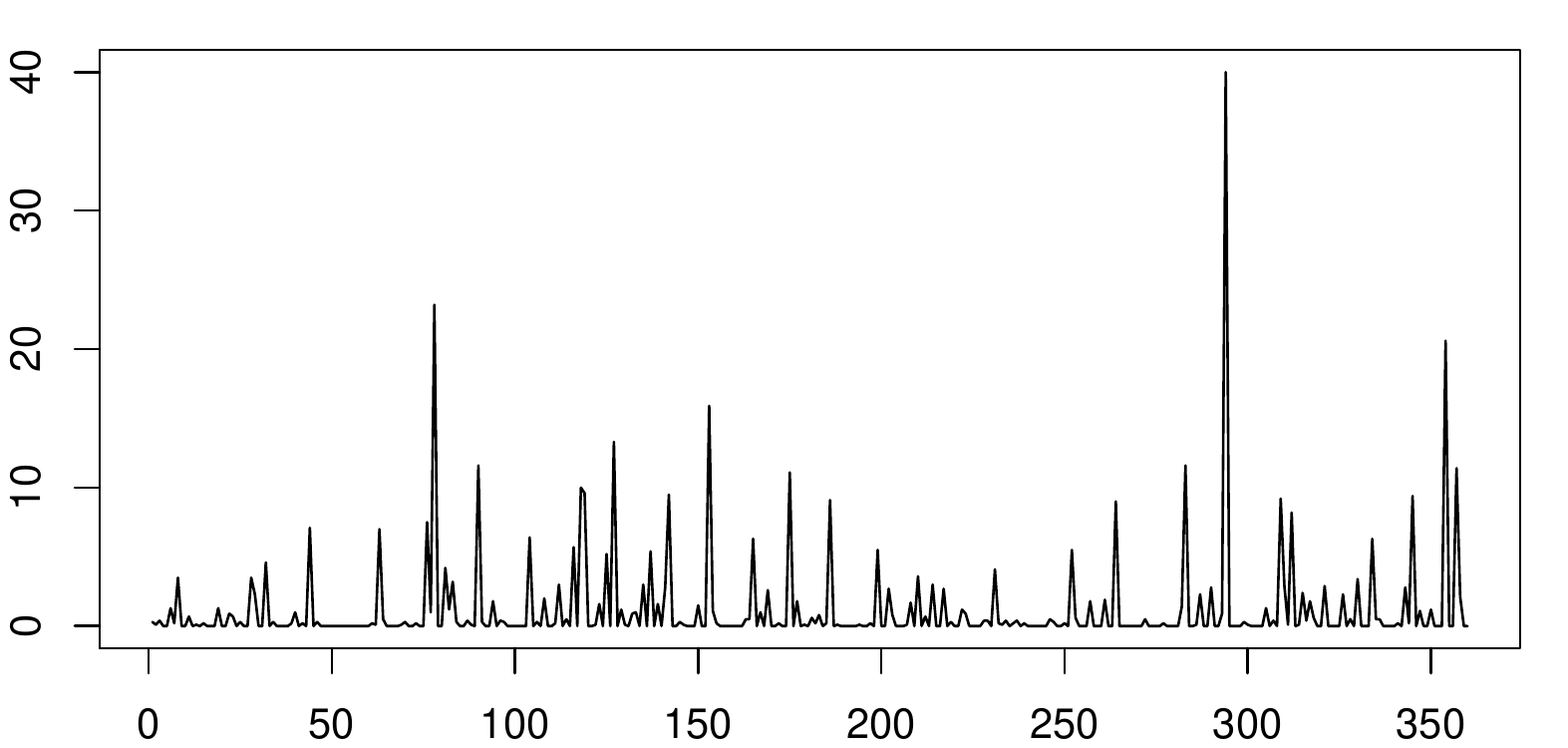}
    \caption{Weather data sampled with monthly frequency. Top part: daily mean temperature (in degrees Celsius). Bottom part: total daily precipitation (in millimeters). Horizontal axis shows the number of months since January 1989.}
    \label{fig:weather}
\end{figure}

As an example of observations from continuous distributions with an atom we consider the data recorded by the Ruzyn\v{e} weather station (Prague, Czech Republic) in the years 1989 to 2018. Namely, for each month in the given period we consider the daily mean temperature (in degrees Celsius) on the 1st day of the month, together with the total daily precipitation (in millimeters) on the same day. Altogether we analyze 360 bivariate observations, see Figure~\ref{fig:weather}. We sample one observation per month to eliminate the autocorrelations in the time series. The observations of total daily precipitation contain nearly 60 \% of 0s, indicating that the corresponding distribution is a mixture of the degenerate distribution in 0 (days with no rain) and a continuous distribution on the positive real numbers (days with rain). The distribution of the daily mean temperature is considered to be continuous. The dataset is available online\footnote{http://portal.chmi.cz/historicka-data/pocasi/denni-data/data-ze-stanic-site-RBCN}.

To assess the independence between the temperature and precipitation we first performed the CDF-based test as described in Section~\ref{se:TS}. The test statistic was computed on a grid of $60 \times 60$ arguments, chosen in an inhomogeneous way, i.e. specified by sample quantiles in each marginal. For the temperature, the quantiles at the probability levels $1/60, 2/60, \ldots$ were used. Due to the presence of the atom, for the precipitation the quantiles were chosen to cover narrower range of probability levels, so that the value of the atom, i.e. 0, is considered only once. The test was based on $9\,999$ permutations and the ERL version of the global envelope test.

\begin{figure}[t]
    \centering
    \includegraphics[width=\textwidth]{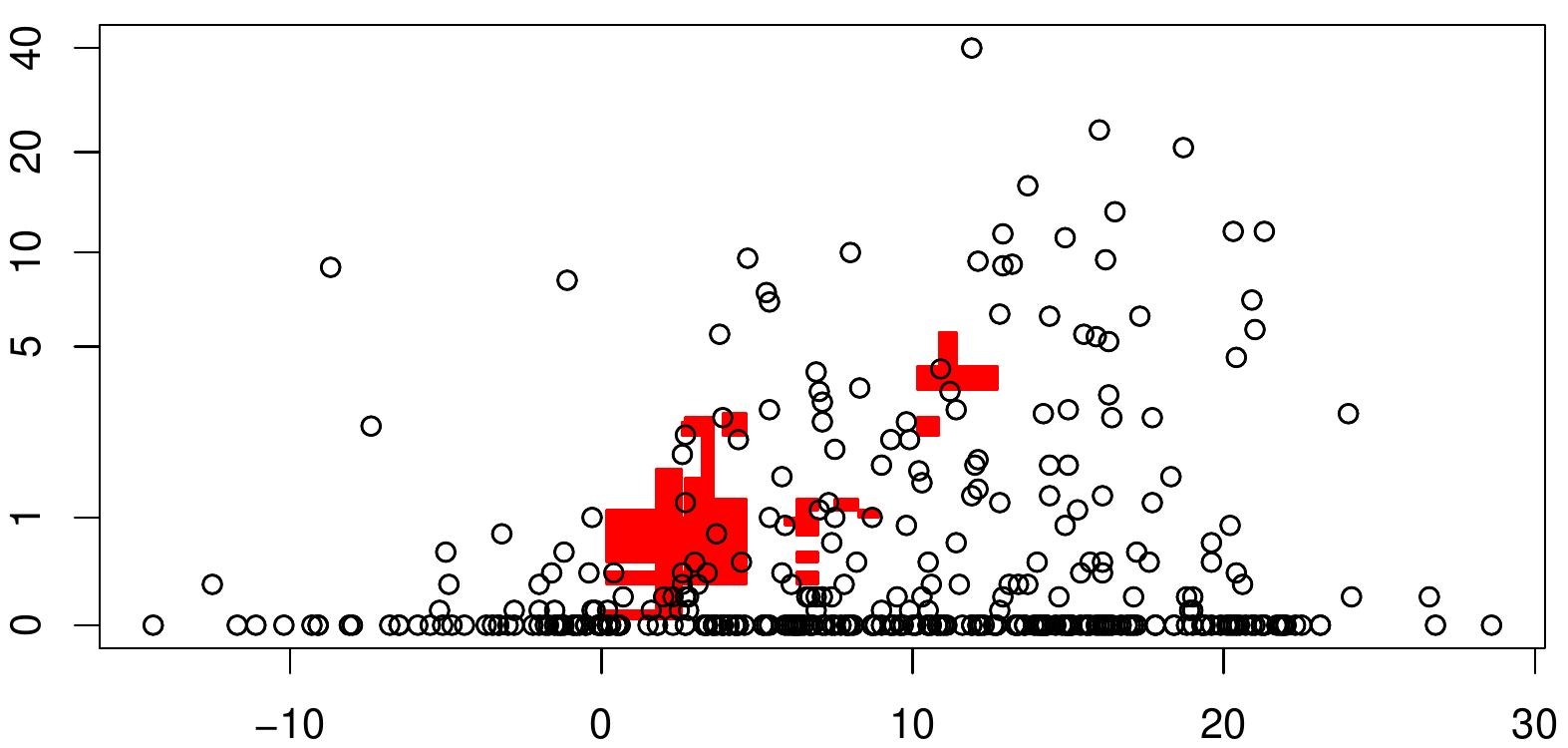}
    \caption{Weather data, CDF-based test of independence. Locations where the observed values exceed the upper envelope are highlighted in red color. Mean daily temperature is given on the horizontal axis, total daily precipitation is given on the vertical axis (note the logarithmic scale of the vertical axis).}
    \label{fig:weather_edf}
\end{figure}

The resulting p-value was 0.001, the regions where the observed values exceeded the upper 95\% envelope are given in Figure~\ref{fig:weather_edf}. Due to the cumulative nature of the test statistic it is rather difficult to interpret the results. Broadly speaking, there appear to be more days with low temperature and low precipitation than what would be expected under independence. This effect accumulates across the region with small values of both variables and finally around the values of 3\degree C (temperature) and 1 mm (precipitation) the accumulated effect is strong enough to be found statistically significant.

\begin{figure}[t]
    \centering
    \includegraphics[width=\textwidth]{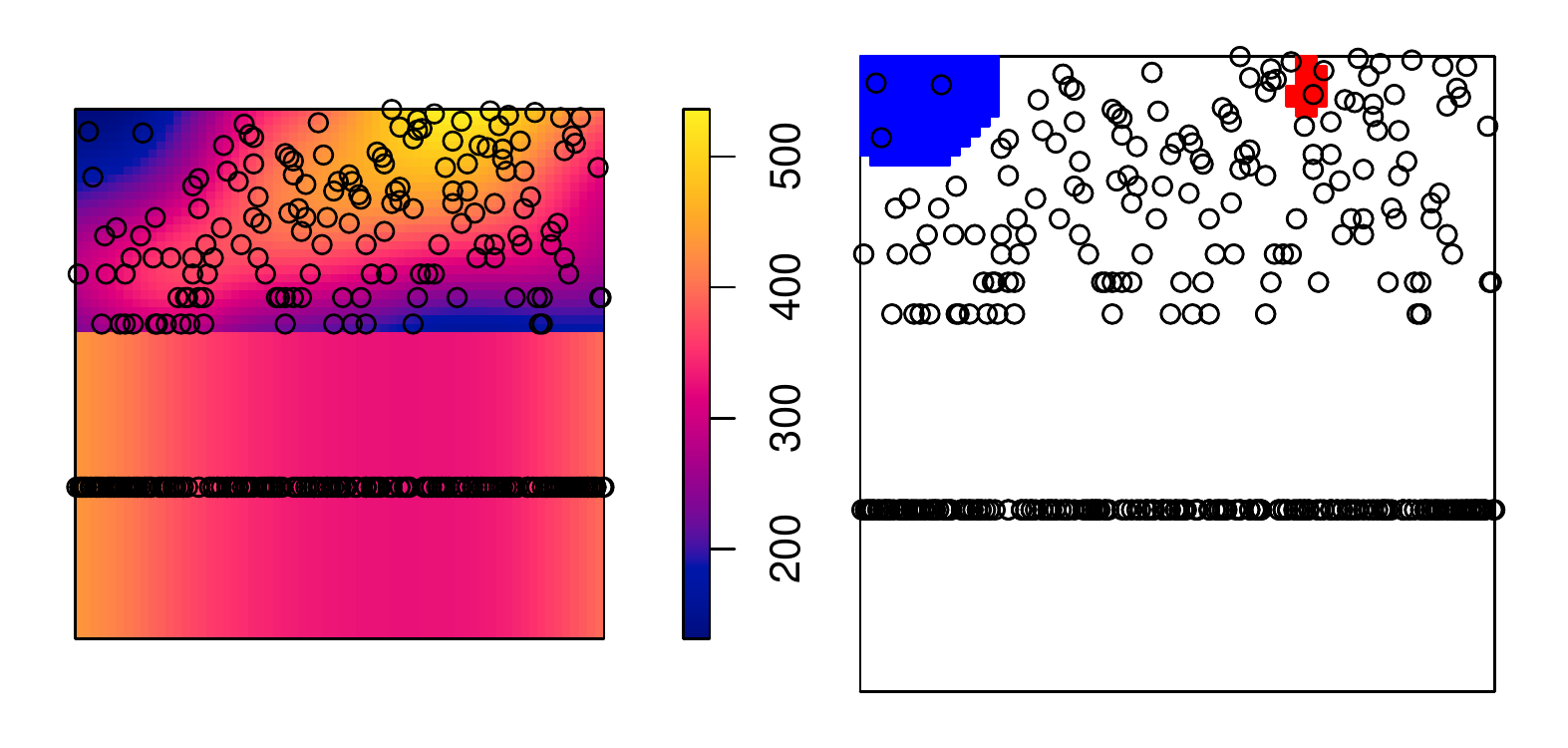}
    \caption{Weather data, QQ-test of independence. Left: QQ representation of the observed data plotted together with the values of the test statistic. Right: graphical output of the global envelope test. Locations where the observed values exceed the upper envelope are highlighted in red color, locations where the observed values are below the lower envelope are highlighted in blue color. Mean daily temperature is given on the horizontal axis, total daily precipitation is given on the vertical axis.}
    \label{fig:weather_qq}
\end{figure}

We further performed the QQ-test as described in Section~\ref{se:TS}. We transformed the observed data using the QQ representation in the unit square and estimated the intensity function on a $64 \times 64$ pixel grid using kernel smoothing, with the Gaussian kernel with standard deviation given by the rule of thumb as $n^{-1/6} / \sqrt{12}$. Again, the presence of the atom requires an adaptation of the procedure. In this case we replaced the appropriate number of rows in the estimated pixel image by values corresponding to the one-dimensional kernel estimate of the one-dimensional intensity function of the points corresponding to the temperature in days with no precipitation, see Figure~\ref{fig:weather_qq}. The fraction of rows to be replaced was chosen to be the (rounded) weight of the atom. In this way all the observations have the same influence in the ERL version of the global envelope test. The estimated values of the one-dimensional intensity function corresponding to the days with no precipitation were also scaled so that the following desirable property of the estimate of the intensity function still holds: integrating the estimated intensity function over the unit square should give the number of observations. Indeed, integrating (numerically) the estimate given in Figure~\ref{fig:weather_qq} gives approx. 360.

The test was based on $9\,999$ permutations and the ERL version of the global envelope test. The resulting p-value was $2 \cdot 10^{-4}$, the graphical output, including the regions where the observed values lie outside the 95\% envelope, is given in Figure~\ref{fig:weather_qq}. The interpretation of the results is easier in this case since the test statistic provides localized information. We observe fewer days with very low temperatures and high precipitation than expected under independence; on the other hand, we observe more days with moderately high temperature and very high precipitation than expected under independence.

For the sake of completeness we also report the p-values obtained from the other tests of independence considered in the simulation study in Secion~\ref{se:SS}: the test based on the Pearson's correlation coefficient (p-value = 0.010), Spearmans's correlation coefficient (0.046), Kendall's correlation coefficient (0.036), the Hoeffding test (0.031), the Genest test (0.032), the DevS test (0.0057), the DevI test (0.0007), the PeaP test (0.0078). All the permutation-based tests were performed using $9\,999$ permutations. The outcomes of all the tests indicate there is a significant dependence between the mean daily temperature and the total daily precipitation. However, unlike the tests discussed above, they do not offer any insight to the reason of rejection of the null hypothesis of independence.

\section{Example for categorical distributions}
\label{se:EC}

As an example of observations from discrete (categorical) distributions we consider the road accidents recorded by the police authorities in Czech Republic in 2018. For each recorded accident we consider two categorical variables: the type of road where the accident occurred and the weather conditions at the time of accident. This is a part of much a bigger dataset, available online\footnote{\url{https://www.policie.cz/clanek/statistika-nehodovosti-900835.aspx}}.

The weather conditions are coded as follows: 1 -- unimpaired, 2 -- fog, 3 -- start of rain or light raining, 4 -- raining, 5 -- snowing, 6 -- icing, 7 -- gusts of wind, 0 -- otherwise impaired. The types of roads are coded as follows: 0 -- highway, 1 -- first class road, 2 -- second class road, 3 -- third class road, 6 -- local road; other types such as field or forest roads are not considered here as the corresponding observations constitute only a marginal part of the dataset. Altogether we analyze $86\,079$ observations. The corresponding contingency table is given in Table~\ref{tab:nehody_GET}.

The classical $\chi^2$-test of independence, with 28 degrees of freedom, in the contingency table results in the p-value smaller than $2.2\cdot 10^{-16}$ (as reported by \texttt{R}). Residuals corresponding to the individual cells are given in Table~\ref{tab:nehody_rezidua}.

The global envelope test can be performed in the contingency table as described in Section \ref{se:TS}. We performed the test with $9\,999$ permutations and significance level 0.05. The output is given in Table~\ref{tab:nehody_GET} (observed values, upper and lower envelope, with indication of significant cells), the resulting p-value was $1 \cdot 10^{-4}$.

The general observation is the following: in good weather conditions (unimpaired, code 1) we observe too many accidents on the local roads (code 6) and too few accidents on higher types of roads (codes 0, 1, 2, 3), compared to what is expected under independence; on the contrary, worse weather conditions (codes 2 to 7) we mostly observe too much accidents on higher types of roads (codes 0, 1, 2, 3) and too few accidents on the local roads (code 6), compared to what is expected under independence. This can be interpreted in the following way: in good weather conditions the accidents are more likely to occur on local roads (with little requirements on turning radii, range of sight etc.), in bad weather conditions the accidents are more likely to occur on higher types of roads (with higher speed limits etc.). {Similar observations can be drawn from the table of residuals in the classical $\chi^2$-test, see Table~\ref{tab:nehody_rezidua}, but without indication which cells are significant.}

\begin{table}[ht]
\centering
\begin{tabular}{r||rrrrrrrr}
  \hline \hline
 & 0 & 1 & 2 & 3 & 4 & 5 & 6 & 7 \\ 
  \hline \hline
0 & 18 & \textcolor{blue}{3273} & 37 & \textcolor{red}{192} & \textcolor{red}{342} & \textcolor{red}{142} & 45 & 4 \\ 
  1 & \textcolor{red}{51} & \textcolor{blue}{12485} & \textcolor{red}{175} & \textcolor{red}{691} & \textcolor{red}{699} & 313 & 163 & \textcolor{red}{41} \\ 
  2 & 33 & \textcolor{blue}{13345} & \textcolor{red}{193} & \textcolor{red}{667} & 590 & \textcolor{red}{467} & \textcolor{red}{236} & \textcolor{red}{40} \\ 
  3 & 32 & \textcolor{blue}{11554} & \textcolor{red}{139} & 489 & \textcolor{blue}{397} & \textcolor{red}{355} & \textcolor{red}{262} & 15 \\ 
  6 & \textcolor{blue}{55} & \textcolor{red}{35675} & \textcolor{blue}{97} & \textcolor{blue}{929} & \textcolor{blue}{966} & \textcolor{blue}{592} & \textcolor{blue}{267} & \textcolor{blue}{13} \\ 
   \hline
  \hline
 & 0 & 1 & 2 & 3 & 4 & 5 & 6 & 7 \\ 
  \hline \hline
0 & 19 & 3653 & 49 & 177 & 177 & 119 & 66 & 14 \\ 
  1 & 49 & 13072 & 139 & 569 & 573 & 369 & 205 & 32 \\ 
  2 & 52 & 13920 & 147 & 604 & 608 & 390 & 214 & 34 \\ 
  3 & 45 & 11850 & 127 & 517 & 519 & 339 & 187 & 30 \\ 
  6 & 107 & 34369 & 325 & 1410 & 1428 & 904 & 487 & 67 \\ 
   \hline
  \hline
 & 0 & 1 & 2 & 3 & 4 & 5 & 6 & 7 \\ 
  \hline \hline
0 & 2 & 3532 & 15 & 106 & 107 & 62 & 27 & 0 \\ 
  1 & 17 & 12858 & 80 & 442 & 442 & 271 & 132 & 8 \\ 
  2 & 19 & 13696 & 88 & 475 & 474 & 288 & 139 & 8 \\ 
  3 & 15 & 11637 & 72 & 395 & 401 & 241 & 117 & 7 \\ 
  6 & 62 & 34082 & 248 & 1248 & 1255 & 769 & 389 & 34 \\ 
   \hline
\end{tabular}
\caption{Permutation-based test of independence in a contingency table for the road accidents dataset, based on $9\,999$ permutations, significance level 0.05. Top part: the observed counts, rows determine the type of road, columns determine the weather conditions. Color coding is used to indicate in which cells the observed count are above the upper envelope (red) or below the lower envelope (blue). Middle part: values corresponding to the upper envelope. Bottom part: values corresponding to the lower envelope.}\label{tab:nehody_GET}
\end{table}

\begin{table}[ht]
\centering
\begin{tabular}{rrrrrrrrr}
  \hline
 & 0 & 1 & 2 & 3 & 4 & 5 & 6 & 7 \\ 
  \hline
0 & 3.05 & -5.36 & 1.24 & 4.42 & 16.93 & 5.76 & -0.12 & -0.57 \\ 
  1 & 3.34 & -4.20 & 6.34 & 8.33 & 8.45 & -0.25 & -0.17 & 4.98 \\ 
  2 & -0.20 & -3.94 & 7.16 & 5.62 & 2.08 & 7.01 & 4.52 & 4.33 \\ 
  3 & 0.54 & -1.75 & 4.07 & 1.52 & -2.96 & 3.98 & 9.18 & -0.57 \\ 
  6 & -3.23 & 7.84 & -11.23 & -11.01 & -10.27 & -8.50 & -8.10 & -5.29 \\ 
   \hline
\end{tabular}
\caption{$\chi^2$-test of independence in a contingency table for the road accidents dataset: residuals corresponding to the individual cells, to be compared with the indicated significant cells in Table~\ref{tab:nehody_GET}.}\label{tab:nehody_rezidua}
\end{table}

\section{Conclusions and discussion}
\label{se:CD}

{In this paper we proposed two model-free tests with graphical interpretation, indicating the reason for possible rejection. This is beneficial in helping the user formulate new models and hypotheses about the observed data. The tests have a broad range of applicability and they can be used both for continuous and discrete distributions, mixtures of those, distributions with heavy tails etc. Specifically, the QQ test can be used for any distribution and the CDF test can be applied to any distribution if the grid of arguments for the test statistic $\hat F(x,y)$ is chosen reasonably by the user}

{As generally accepted, no test of independence is uniformly better than others. For this reason we have not compared our proposed tests to a broad range of tests currently available but rather we have chosen a set of benchmark tests to compare with. The main benefit of our tests lies in the graphical interpretation of the outcome of the tests, not in outperforming all others.}

{The simulation study presented above provides insight into the performance of the tests and indicates in which settings the CDF test performs well (departures from independence appearing over the whole range of data values) and in which the QQ test performs well (departures from independence localized). Hence the two tests are complementary in a certain sense and the user can choose which test to use based on prior assumptions about the given dataset.}


%
%

\bibliographystyle{spbasic}      
\bibliography{Tomas_bibfile}   

\end{document}